\documentclass[showpacs,preprintnumbers,aps]{revtex4}
\usepackage{graphicx}
\usepackage{dcolumn}
\usepackage{bm}

\begin{document}

\title{Quantum Bose Josephson Junction with binary mixtures of BECs}
\author{Adele Naddeo}
\email{naddeo@sa.infn.it}
\author{Roberta Citro}
\email{citro@sa.infn.it}

\affiliation{Dipartimento di Fisica "E. R. Caianiello",
Universit\'{a} degli Studi di Salerno and CNISM, Unit\'{a} di
Ricerca di Salerno, Via Ponte Don Melillo, 84084 Fisciano (SA),
Italy}

\date{\today}

\begin{abstract}
We study the quantum behaviour of a binary mixture of
Bose-Einstein condensates (BEC) in a double-well potential
starting from a two-mode Bose-Hubbard Hamiltonian. We focus on the
small tunneling amplitude regime and apply perturbation theory up
to second order. Analytical expressions for the energy eigenvalues
and eigenstates are obtained. Then the quantum evolution of the
number difference of bosons between the two potential wells is
fully investigated for two different initial conditions:
completely localized states and coherent spin states. In the first
case both the short and the long time dynamics is studied and a
rich behaviour is found, ranging from small amplitude oscillations
and collapses and revivals to coherent tunneling. In the second
case the short-time scale evolution of number difference is
determined and a more irregular dynamics is evidenced.
Finally, the formation of Schroedinger cat states is considered
and shown to affect the momentum distribution.
\end{abstract}

 \maketitle


\section{Introduction}

The experimental discovery of Bose-Einstein condensation
\cite{bec1} in dilute systems of trapped alkali-metal atoms, such
as rubidium ($Rb$), lithium ($Li$), sodium ($Na$) and ytterbium
($Yb$), has spurred a renewed interest into the investigation of
macroscopic quantum phenomena and interference effects, allowing
for a deeper understanding of the conceptual foundations of
quantum mechanics \cite{qm1}. This fascinating research area has
been growing up thanks to the high degree of experimental
manipulation and control \cite{bec2}. Interference between
condensates released in a potential with a barrier was first
observed in 1997 \cite{bec3} and that paved the way for further
investigations on the problem of Bose condensates in a double well
potential. Then Josephson oscillations have been observed in one
dimensional optical potential arrays \cite{bec4}. A single bosonic
Josephson junction was produced for the first time in 2005 with
$Rb$ atoms and its dynamics was experimentally investigated both
within tunneling as well as self-trapping regime
\cite{bec5}\cite{gati1}\cite{levy}. More recently, mixtures of
$^{85}Rb$ and $^{87}Rb$ atoms have been produced and
experimentally investigated \cite{bec6} as well, whose
intraspecies scattering lengths could be tunable via magnetic and
optical Feshbach resonances. Furthermore the realization of
heteronuclear mixtures of $^{87}Rb$ and $^{41}K$ atoms with
tunable interspecies interactions \cite{bec7} paved the way to the
exploration of double species Mott insulators and, in general, of
the quantum phase diagram of two species Bose-Hubbard model
\cite{2bh1}. The interplay between the interspecies and
intraspecies scattering produces deep consequences on the
properties of the condensates, such as the density profile
\cite{bec8} and the collective excitations \cite{bec9}. However,
the wide tunability of such interactions makes a BEC mixture a
very interesting subject of investigation, both from experimental
and theoretical side as a mean of studying new macroscopic quantum
tunneling phenomena as well as the interplay between quantum
coherence and nonlinearity. Indeed novel and richer behaviours are
expected in such a multicomponent BEC.

On the theoretical side, a bosonic Josephson junction with a
single species of BEC has been widely investigated by means of a
two-mode approximation
\cite{milburn}\cite{smerzi1}\cite{ananikian1}, within the
classical as well as the quantum regime. In the classical regime,
characterized by large particle numbers and weak repulsive
interactions, the Gross-Pitaevskii equation provides a reliable
description. Within the two mode approximation it reduces to two
generalized Josephson equations which describe the time evolution
of the relative phase and the population imbalance between the
wells \cite{smerzi1} and differ from their superconducting
counterpart \cite{barone1} by the presence of a nonlinear term
which couples the variables. Because of such a term, a bosonic
Josephson junction exhibit a variety of novel phenomena which
range from $\pi $-oscillations to macroscopic quantum
self-trapping (MQST) \cite{smerzi1}. While the $\pi
$-oscillations, as well the usual Josephosn ones, deal with a
symmetric oscillation of the condensate about the two wells, the
MQST phenomenon is characterized by a broken symmetry phase with a
population imbalance between the wells. In the quantum regime,
characterized by smaller values of the particle number and strong
interactions, an increasing of phase fluctuations is observed
together with the suppression of number fluctuations. Furthermore
the time evolution is characterized by phase collapse and revival
\cite{cr1}. The quantum behaviour of bosonic Josephson junctions
has been deeply investigated by means of the usual quantum phase
model \cite{zapata1}\cite{pitaevskii2}\cite {smerzi2} as well as
by starting from a two-mode Bose-Hubbard Hamiltonian
\cite{bishop1}\cite{bishop2}\cite{minguzzi1}. In this context the
phase coherence of the junction has been characterized by studying
the momentum distribution \cite{pitaevskii1}\cite{pitaevskii2}.
The generation and detection of Schroedinger cat states has been
investigated as well; indeed the presence of such a kind of states
reflects in the strong reduction of the momentum-distribution
contrast \cite{minguzzi1}\cite{minguzzi2}.

More recently such a theoretical analysis has been successfully
extended to a binary mixture of BECs in a double well potential
\cite{mix2}\cite{mix1} \cite{mix3}\cite{mix4}. The semiclassical
regime in which the fluctuations around the mean values are small
has been deeply investigated and found to be described by two
coupled Gross-Pitaevskii equations. By means of a two-mode
approximation such equations can be cast in the form of four
coupled nonlinear ordinary differential equations for the
population imbalance and the relative phase of each species. The
solution results in a richer tunneling dynamics. In particular,
two different MQST states with broken symmetry have been found
\cite{mix3}, where the two species localize in the two different
wells giving rise to a phase separation or coexist in the same
well respectively.  Indeed, upon a variation of some parameters or
initial conditions, the phase-separated MQST states evolve towards
a symmetry-restoring phase where the two components swap places
between the two wells, so avoiding each other. Recently, the
coherent dynamics of a two species BEC in a double well has been
analyzed as well focussing on the case where the two species are
two hyperfine states of the same alkali metal \cite{mix5}.

In this paper we study the quantum behaviour of a binary mixture
of Bose-Einstein condensates (BEC) in a double-well potential
starting from a two-mode Bose-Hubbard Hamiltonian. We analyze in
detail the small tunneling amplitude regime where number
fluctuations are suppressed and a Mott-insulator behaviour is
established. We perform a perturbative calculation up to second
order in the tunneling amplitude and study the stationary states
and the dynamics of the two species bosonic Josephson junction.
Finally, the dynamical generation of Schroedinger cat states is
investigated starting from an initial coherent spin state and
shown to affect the time-dependent population imbalance and
momentum distribution\cite {pitaevskii1}\cite{pitaevskii2}. We
focus on the contrast in the momentum distribution between the two
wells and show how it vanishes for a two-component cat state. That
could be interesting in view of the experimental realization of
macroscopic superpositions of quantum states
\cite{minguzzi2}\cite{smerzi3}.

The paper is organized as follows. In Section 2 we introduce our
model Hamiltonian within the two-mode approximation and define the
various parameters. Then we adopt the angular momentum
representation and focus on the small tunneling amplitude regime.
In Section 3 we apply perturbation theory in the tunneling
amplitude to our Hamiltonian and find analytical expressions for
the energy eigenvalues and eigenstates up to second order. Section
4 and 5 are devoted to the study of the quantum evolution of the
number difference of bosons between the two wells in
correspondence of two different initial conditions: completely
localized states and coherent spin states. In the first case both
the short and the long time dynamics is studied and a rich
behaviour is evidenced, ranging from small amplitude oscillations
and collapses and revivals to coherent tunneling. In the second
case the short-time scale evolution of number difference is
determined and a more irregular dynamics is evidenced, with suppression of the dominant
frequency when the number of bosons increase. Then, Schroedinger
cat states are shown to generate as a result of the time-evolution
of an initial coherent state when the tunneling between the two
wells is suppressed, and their influence on the contrast in the
momentum distribution is studied. Finally, in Section 6 some
conclusions and outlooks of this work are presented.

\section{The model}

A binary mixture of Bose-Einstein condensates
\cite{mix1}\cite{mix3} loaded in a double-well potential is
described by the general many-body Hamiltonian:
\begin{equation}
H=H_{a}+H_{b}+H_{ab}  \label{m1}
\end{equation}
where
\begin{equation}
H_{a}=\int d\overrightarrow{r}\left( -\frac{\hbar
^{2}}{2m_{a}}\psi
_{a}^{+}\nabla ^{2}\psi _{a}+\psi _{a}^{+}V_{a}\left( \overrightarrow{r}%
\right) \psi _{a}\right) +\frac{1}{2}\int \int d\overrightarrow{r}d%
\overrightarrow{r}^{\prime }\psi _{a}^{+}\left(
\overrightarrow{r}\right) \psi _{a}^{+}\left(
\overrightarrow{r}^{\prime }\right) U_{aa}\left(
\overrightarrow{r}-\overrightarrow{r}^{\prime }\right) \psi
_{a}\left(
\overrightarrow{r}^{\prime }\right) \psi _{a}\left( \overrightarrow{r}%
\right) ,  \label{m2}
\end{equation}
\begin{equation}
H_{b}=\int d\overrightarrow{r}\left( -\frac{\hbar
^{2}}{2m_{b}}\psi
_{b}^{+}\nabla ^{2}\psi _{b}+\psi _{b}^{+}V_{b}\left( \overrightarrow{r}%
\right) \psi _{b}\right) +\frac{1}{2}\int \int d\overrightarrow{r}d%
\overrightarrow{r}^{\prime }\psi _{b}^{+}\left(
\overrightarrow{r}\right) \psi _{b}^{+}\left(
\overrightarrow{r}^{\prime }\right) U_{bb}\left(
\overrightarrow{r}-\overrightarrow{r}^{\prime }\right) \psi
_{b}\left(
\overrightarrow{r}^{\prime }\right) \psi _{b}\left( \overrightarrow{r}%
\right)  \label{m3}
\end{equation}
are the Hamiltonians for bosons of species $a$ and $b$
respectively and
\begin{equation}
H_{ab}=\int \int d\overrightarrow{r}d\overrightarrow{r}^{\prime
}\psi
_{a}^{+}\left( \overrightarrow{r}\right) \psi _{b}^{+}\left( \overrightarrow{%
r}^{\prime }\right) U_{ab}\left( \overrightarrow{r}-\overrightarrow{r}%
^{\prime }\right) \psi _{a}\left( \overrightarrow{r}^{\prime
}\right) \psi _{b}\left( \overrightarrow{r}\right) ,  \label{m4}
\end{equation}
is the interaction term between bosons of different species. For
dilute mixtures one can replace the interaction potentials
$U_{aa}$, $U_{bb}$ and $U_{ab}$ with the effective contact
interactions:
\begin{equation}
\begin{array}{ccc}
U_{aa}\left( \overrightarrow{r}-\overrightarrow{r}^{\prime
}\right) =g_{aa}\delta \left(
\overrightarrow{r}-\overrightarrow{r}^{\prime }\right) , &
U_{bb}\left( \overrightarrow{r}-\overrightarrow{r}^{\prime
}\right) =g_{bb}\delta \left( \overrightarrow{r}-\overrightarrow{r}%
^{\prime }\right) , & U_{ab}\left( \overrightarrow{r}-\overrightarrow{r}%
^{\prime }\right) =g_{ab}\delta \left( \overrightarrow{r}-\overrightarrow{%
r}^{\prime }\right)
\end{array}
,  \label{m5}
\end{equation}
where $g_{aa}=\frac{4\pi \hbar ^{2}a_{aa}}{m_{a}}$ and
$g_{bb}=\frac{4\pi \hbar ^{2}a_{bb}}{m_{b}}$ are the intraspecies
coupling constants of the species $a$ and $b$ respectively,
$m_{a}$ and $m_{b}$ being the atomic masses
and $a_{aa}$, $a_{bb}$ the $s$-wave scattering lengths; furthermore $g_{ab}=%
\frac{2\pi \hbar ^{2}a_{ab}}{m_{ab}}$ is the interspecies coupling
constant,
where $m_{ab}=\frac{m_{a}m_{b}}{m_{a}+m_{b}}$ is the reduced mass and $%
a_{ab} $ is the associated $s$-wave scattering length. In this way
the Hamiltonian (\ref{m1})-(\ref{m4}) can be rewritten as:
\begin{equation}
H_{i}=\int d\overrightarrow{r}\left( -\frac{\hbar
^{2}}{2m_{i}}\psi
_{i}^{+}\nabla ^{2}\psi _{i}+\psi _{i}^{+}V_{i}\left( \overrightarrow{r}%
\right) \psi _{i}\right) +\frac{g_{ii}}{2}\int
d\overrightarrow{r}\psi _{i}^{+}\psi _{i}^{+}\psi _{i}\psi
_{i};\text{ \ \ \ }i=a,b  \label{m6}
\end{equation}
\begin{equation}
H_{ab}=g_{ab}\int d\overrightarrow{r}\psi _{a}^{+}\psi
_{b}^{+}\psi _{a}\psi _{b}.  \label{m7}
\end{equation}
Here $V_{i}\left( \overrightarrow{r}\right) $ is the double well
trapping
potential and, in the following, we assume $V_{a}\left( \overrightarrow{r}%
\right) =V_{b}\left( \overrightarrow{r}\right) =V\left( \overrightarrow{r}%
\right) $; $\psi _{i}^{+}\left( \overrightarrow{r}\right) ,$ $\psi
_{i}\left( \overrightarrow{r}\right) $, $i=a,b$ are the bosonic
creation and annihilation operators for the two species, which
satisfy the commutation rules:
\begin{eqnarray}
\left[ \psi _{i}\left( \overrightarrow{r}\right) ,\psi _{j}\left(
\overrightarrow{r}^{\prime }\right) \right] &=&\left[ \psi
_{i}^{+}\left( \overrightarrow{r}\right) ,\psi _{j}^{+}\left(
\overrightarrow{r}^{\prime
}\right) \right] =0,  \label{m8} \\
\left[ \psi _{i}\left( \overrightarrow{r}\right) ,\psi
_{j}^{+}\left( \overrightarrow{r}^{\prime }\right) \right]
&=&\delta _{ij}\delta \left(
\overrightarrow{r}-\overrightarrow{r}^{\prime }\right) ,\text{ \ \ \ \ }%
i,j=a,b,  \label{m9}
\end{eqnarray}
and the normalization conditions:
\begin{equation}
\int d\overrightarrow{r}\left| \psi _{i}\left(
\overrightarrow{r}\right) \right| ^{2}=N_{i};\text{ \ \ \ }i=a,b,
\label{m10}
\end{equation}
$N_{i}$, $i=a,b$ being the number of atoms of species $a$ and $b$
respectively. The total number of atoms of the mixture is
$N=N_{a}+N_{b}$.

Now a weak link between the two wells produces a small energy
splitting between the mean-field ground state and the first
excited state of the double well potential and that allows us to
reduce the dimension of the Hilbert space of the initial many-body
problem. Indeed for low energy excitations and low temperatures it
is possible to consider only such two states and neglect the
contribution from the higher ones, the so called two-mode
approximation \cite{milburn} \cite{smerzi1}\cite{ananikian1}. In
this way, by taking into account for each of the two species $a$
and $b$ the mean-field ground states $\phi
_{g}^{a}$, $\phi _{g}^{b}$ and the mean-field excited states $\phi _{e}^{a}$%
, $\phi _{e}^{b}$, the wave functions $\psi _{i}$, $i=a,b$, can be
rewritten as:
\begin{equation}
\begin{array}{c}
\psi _{a}=a_{g}\phi _{g}^{a}+a_{e}\phi _{e}^{a} \\
\psi _{b}=b_{g}\phi _{g}^{b}+b_{e}\phi _{e}^{b}
\end{array}
,  \label{m11}
\end{equation}
where $\int d\overrightarrow{r}\left| \phi _{g\left( e\right)
}^{i}\right|
^{2}=1$, $i=a,b$, and $a_{g}^{+}$, $b_{g}^{+}$ and $a_{e}^{+}$, $b_{e}^{+}$ (%
$a_{g}$, $b_{g}$ and $a_{e}$, $b_{e}$) are the creation
(annihilation) operators for a particle of the species $a$, $b$ in
the ground and the excited state respectively. They satisfy the
usual bosonic commutation relations $\left[ a_{i},a_{j}^{+}\right]
=\left[ b_{i},b_{j}^{+}\right] =\delta _{ij}$. Furthermore $\phi
_{g,e}^{i}$, $i=a,b$ are assumed real for simplicity and such that
$\left\langle \left( \phi _{g}^{i}\right) ^{3}\phi
_{e}^{j}\right\rangle =\left\langle \left( \phi _{e}^{i}\right)
^{3}\phi _{g}^{j}\right\rangle =0$, which simplifies the
calculations. Let us change the basis and switch to the atom
number states in such a way that the expectation value of the
population of the left and right well can be defined. The new
annihilation operators are $a_{L}=\frac{1}{\sqrt{2}}\left(
a_{g}+a_{e}\right) $, $a_{R}=\frac{1}{\sqrt{2}}\left(
a_{g}-a_{e}\right) $
and $b_{L}=\frac{1}{\sqrt{2}}\left( b_{g}+b_{e}\right) $, $b_{R}=\frac{1}{%
\sqrt{2}}\left( b_{g}-b_{e}\right) $ for the species $a$ and $b$
respectively, so that the wave functions (\ref{m11}) become:
\begin{equation}
\begin{array}{c}
\psi _{a}=\frac{1}{\sqrt{2}}a_{L}\left( \phi _{g}^{a}+\phi _{e}^{a}\right) +%
\frac{1}{\sqrt{2}}a_{R}\left( \phi _{g}^{a}-\phi _{e}^{a}\right)  \\
\psi _{b}=\frac{1}{\sqrt{2}}b_{L}\left( \phi _{g}^{b}+\phi _{e}^{b}\right) +%
\frac{1}{\sqrt{2}}b_{R}\left( \phi _{g}^{b}-\phi _{e}^{b}\right)
\end{array}
.  \label{m12}
\end{equation}
By substituting Equations (\ref{m12}) into the Hamiltonian
(\ref{m6})-(\ref{m7}), after some algebra we obtain its second
quantized version within the two-mode approximation:
\begin{eqnarray}
H &=&\frac{E_{c}^{a}}{8}\left( a_{R}^{+}a_{R}-a_{L}^{+}a_{L}\right) ^{2}-%
\frac{\overline{E}_{J}^{a}}{N_{a}}\left(
a_{R}^{+}a_{L}+a_{L}^{+}a_{R}\right) +\delta E^{a}\left(
a_{R}^{+}a_{L}+a_{L}^{+}a_{R}\right)
^{2}+\frac{E_{c}^{b}}{8}\left(
b_{R}^{+}b_{R}-b_{L}^{+}b_{L}\right) ^{2}   \nonumber \\
&&-\frac{\overline{E}_{J}^{b}}{N_{b}}\left(
b_{R}^{+}b_{L}+b_{L}^{+}b_{R}\right) +\delta E^{b}\left(
b_{R}^{+}b_{L}+b_{L}^{+}b_{R}\right) ^{2}+\frac{1}{4}\Lambda
_{ab}\left( a_{L}^{+}a_{L}-a_{R}^{+}a_{R}\right) \left(
b_{L}^{+}b_{L}-b_{R}^{+}b_{R}\right) +   \nonumber \\
&&-\frac{1}{4}\left( a_{R}^{+}a_{L}+a_{L}^{+}a_{R}\right) \left(
b_{R}^{+}b_{L}+b_{L}^{+}b_{R}\right) \left( \kappa
_{e,g}^{ab}+\kappa
_{g,e}^{ab}-\kappa _{g,g}^{ab}-\kappa _{e,e}^{ab}\right) +\frac{1}{2}%
N_{a}\left( E_{g}^{a}+E_{e}^{a}\right) +   \label{m13} \\
&&+\frac{1}{4}N_{a}\left( N_{a}-2\right) \left( \kappa
_{g,g}^{a}+\kappa _{e,e}^{a}\right) +\left(
N_{aL}^{2}+N_{aR}^{2}-N_{a}\right) \kappa
_{g,e}^{a}+\frac{1}{2}N_{b}\left( E_{g}^{b}+E_{e}^{b}\right) +   \nonumber \\
&&+\frac{1}{4}N_{b}\left( N_{b}-2\right) \left( \kappa
_{g,g}^{b}+\kappa _{e,e}^{b}\right) +\left(
N_{bL}^{2}+N_{bR}^{2}-N_{b}\right) \kappa
_{g,e}^{b}+\frac{1}{4}N_{a}N_{b}\left( \kappa _{e,g}^{ab}+\kappa
_{g,e}^{ab}+\kappa _{g,g}^{ab}+\kappa _{e,e}^{ab}\right) ,
\nonumber
\end{eqnarray}
where $N_{i}=N_{iL}+N_{iR}$, $i=a,b$, is the number of atoms of
species $a$ and $b$ respectively, expressed as a sum of numbers of
atoms in the left and right well. The parameters are defined as
follows:
\begin{equation}
E_{g}^{i}=\int d\overrightarrow{r}\left( -\frac{\hbar
^{2}}{2m_{i}}\phi _{g}^{i}\nabla ^{2}\phi _{g}^{i}+\phi
_{g}^{i}V\phi _{g}^{i}\right) ;\text{ \ \ \ }i=a,b  \label{m14}
\end{equation}
\begin{equation}
E_{e}^{i}=\int d\overrightarrow{r}\left( -\frac{\hbar
^{2}}{2m_{i}}\phi _{e}^{i}\nabla ^{2}\phi _{e}^{i}+\phi
_{e}^{i}V\phi _{e}^{i}\right) ;\text{ \ \ \ }i=a,b  \label{m15}
\end{equation}
\begin{equation}
\kappa _{i,j}^{a}=\frac{g_{aa}}{2}\int d\overrightarrow{r}\left|
\phi _{i}^{a}\right| ^{2}\left| \phi _{j}^{a}\right| ^{2};\text{ \
\ \ }i,j=g,e \label{m16}
\end{equation}
\begin{equation}
\kappa _{i,j}^{b}=\frac{g_{bb}}{2}\int d\overrightarrow{r}\left|
\phi _{i}^{b}\right| ^{2}\left| \phi _{j}^{b}\right| ^{2};\text{ \
\ \ }i,j=g,e \label{m17}
\end{equation}
\begin{equation}
\kappa _{i,j}^{ab}=g_{ab}\int d\overrightarrow{r}\left| \phi
_{i}^{a}\right| ^{2}\left| \phi _{j}^{b}\right| ^{2};\text{ \ \ \
}i,j=g,e  \label{m18}
\end{equation}
\begin{equation}
\Lambda _{ab}=4g_{ab}\int d\overrightarrow{r}\phi _{g}^{a}\phi
_{e}^{a}\phi _{g}^{b}\phi _{e}^{b}; \label{m19}
\end{equation}
\begin{equation}
E_{c}^{i}=4\kappa _{g,e}^{i};\text{ \ \ \ }i=a,b  \label{m20}
\end{equation}
\begin{equation}
\delta E^{i}=\frac{\kappa _{g,g}^{i}+\kappa _{e,e}^{i}-2\kappa _{g,e}^{i}}{4}%
;\text{ \ \ \ }i=a,b  \label{m21}
\end{equation}
\begin{equation}
\overline{E}_{J}^{i}=\frac{N_{i}}{2}\left( E_{e}^{i}-E_{g}^{i}\right) +\frac{%
N_{i}}{2}\left[ \left( N_{i}-1\right) \left( \kappa
_{g,g}^{i}-\kappa _{e,e}^{i}\right) +\frac{N_{j}}{2}\left( \kappa
_{e,e}^{ab}-\kappa _{g,g}^{ab}+\kappa _{g,e}^{ab}-\kappa
_{e,g}^{ab}\right) \right] ;\text{ \ \ \ }i,j=a,b;\text{ \ \ \
}i\neq j.  \label{m22}
\end{equation}
In Eq. (\ref{m13}), the terms proportional to
$\overline{E}_{J}^{i}$, $i=a,b$, describe tunneling of particles
of species $a$ and $b$ from one to the other well while the terms
proportional to $E_{c}^{i}$, $i=a,b$, deal with the local
interaction within the two wells and the terms proportional to
$\delta E^{i}$ correspond to additional two-particle processes.
Finally the terms proportional to $\Lambda_{ab}$ and $\kappa
_{i,j}^{ab}$ couple the two species and then various constant
terms follow, which we will drop for simplicity.

In this paper we focus on the small tunneling amplitude regime
where number fluctuations are suppressed and a Mott-insulator
behaviour is established, so it is convenient to introduce the
angular momentum representation for the species $a$ and $b$ as
follows:
\begin{equation}
\begin{array}{ccc}
J_{x}^{a}=\frac{1}{2}\left( a_{R}^{+}a_{L}+a_{L}^{+}a_{R}\right) ,
& J_{y}^{a}=\frac{i}{2}\left( a_{R}^{+}a_{L}-a_{L}^{+}a_{R}\right)
, &
J_{z}^{a}=\frac{1}{2}\left( a_{R}^{+}a_{R}-a_{L}^{+}a_{L}\right) , \\
J_{x}^{b}=\frac{1}{2}\left( b_{R}^{+}b_{L}+b_{L}^{+}b_{R}\right) ,
&
J_{y}^{b}=\frac{i}{2}\left( b_{R}^{+}b_{L}-b_{L}^{+}b_{R}\right) & J_{z}^{b}=%
\frac{1}{2}\left( b_{R}^{+}b_{R}-b_{L}^{+}b_{L}\right) ,
\end{array}
\label{m23}
\end{equation}
where the operators $J_{i}^{a}$, $J_{i}^{b}$, $i=x,y,z$, obey to
the usual angular momentum algebra and the following relations
hold:
\begin{equation}
\begin{array}{cc}
\left( J^{a}\right) ^{2}=\frac{N_{a}}{2}\left(
\frac{N_{a}}{2}+1\right) , & \left( J^{b}\right)
^{2}=\frac{N_{b}}{2}\left( \frac{N_{b}}{2}+1\right) .
\end{array}
\label{m23a}
\end{equation}
In particular, the components $J_{z}^{i}=\frac{1}{2}\left(
N_{iR}-N_{iL}\right) $, $i=a,b$ give the difference of the number
of bosons of the species $i$, $N_{iL}$ and $N_{iR}$, occupying the
two minima of the double well potential, i. e. the population
imbalances, which are experimentally observable quantities. Thus
Hamiltonian (\ref{m13}) can be cast in the following form:
\begin{eqnarray}
H &=&\frac{E_{c}^{a}}{2}\left( J_{z}^{a}\right) ^{2}-2\frac{\overline{E}%
_{J}^{a}}{N_{a}}J_{x}^{a}+4\delta E^{a}\left( J_{x}^{a}\right) ^{2}+\frac{%
E_{c}^{b}}{2}\left( J_{z}^{b}\right) ^{2}-2\frac{\overline{E}_{J}^{b}}{N_{b}}%
J_{x}^{b}+4\delta E^{b}\left( J_{x}^{b}\right) ^{2}+  \nonumber \\
&&+\Lambda _{ab}J_{z}^{a}J_{z}^{b}-J_{x}^{a}J_{x}^{b}\left( \kappa
_{e,g}^{ab}+\kappa _{g,e}^{ab}-\kappa _{g,g}^{ab}-\kappa
_{e,e}^{ab}\right) , \label{m24}
\end{eqnarray}
where the constant terms have been dropped. Let us now simplify
the notation by introducing the following parameters:
\begin{equation}
\begin{array}{c}
\begin{array}{ccc}
\Lambda _{a}=E_{c}^{a}, & C_{a}=4\delta E^{a}, & K_{a}=2\frac{\overline{E}%
_{J}^{a}}{N_{a}}, \\
\Lambda _{b}=E_{c}^{b}, & C_{b}=4\delta E^{b}, & K_{b}=2\frac{\overline{E}%
_{J}^{b}}{N_{b}},
\end{array}
\\
D_{ab}=\kappa _{e,g}^{ab}+\kappa _{g,e}^{ab}-\kappa
_{g,g}^{ab}-\kappa _{e,e}^{ab}
\end{array}
\label{m25}
\end{equation}
and rewrite the Hamiltonian (\ref{m24}) as:
\begin{eqnarray}
H &=&\frac{1}{2}\Lambda _{a}\left( J_{z}^{a}\right)
^{2}-K_{a}J_{x}^{a}+C_{a}\left( J_{x}^{a}\right)
^{2}+\frac{1}{2}\Lambda _{b}\left( J_{z}^{b}\right)
^{2}-K_{b}J_{x}^{b}+C_{b}\left( J_{x}^{b}\right)
^{2}+  \nonumber \\
&&+\Lambda _{ab}J_{z}^{a}J_{z}^{b}-D_{ab}J_{x}^{a}J_{x}^{b}.
\label{m26}
\end{eqnarray}
Within the experimental parameters range it is possible to show that $%
C_{i}\ll \Lambda _{i},K_{i}$, $i=a,b$, and $D_{ab}\ll \Lambda
_{ab}$ \cite
{gati1}\cite{mix3}, then in the following we put $C_{a}=C_{b}=0$ and $%
D_{ab}=0$, which corresponds to neglecting the spatial overlap
integrals between the localized modes in the two wells. In this
way the binary mixture of BECs within two-mode approximation maps
to two Ising-type spin model in a transverse magnetic field.

In the following we will focus on the symmetric case $\Lambda
_{a}=\Lambda _{b}=\Lambda $ and $K_{a}=K_{b}=K$ because it allows
us to perform analytical calculations while capturing many
relevant phenomena characterizing the physics of the system. So
the model Hamiltonian (\ref{m24}) becomes:
\begin{equation}
H=H_{0}+H_{I},  \label{m27}
\end{equation}
\begin{equation}
H_{0}=\frac{1}{2}\Lambda \left( J_{z}^{a}\right)
^{2}+\frac{1}{2}\Lambda \left( J_{z}^{b}\right) ^{2}+\Lambda
_{ab}J_{z}^{a}J_{z}^{b},  \label{m28}
\end{equation}
\begin{equation}
H_{I}=-K\left( J_{x}^{a}+J_{x}^{b}\right) ,  \label{m29}
\end{equation}
where, in the small tunneling amplitude regime, $H_{I}$ is
considered as a perturbation. The total Hamiltonian commutes with
$\left( J^{a}\right) ^{2}$ and $\left( J^{b}\right) ^{2}$, which
leads to the conservation of total
angular momentum with quantum numbers $j_{a}=\frac{N_{a}}{2}$ and $j_{b}=%
\frac{N_{b}}{2}$ respectively. So the whole Hilbert space has
finite dimension, equal to $\left( 2j_{a}+1\right) \otimes \left(
2j_{b}+1\right) =\left( N_{a}+1\right) \otimes \left(
N_{b}+1\right) $, thus it depends on the number of bosons of the
species $a$ and $b$ respectively. The whole basis $\left\{ \left|
m_{a}\right\rangle ,\left| m_{b}\right\rangle \right\} $ is given
by the eigenvectors of $J_{z}^{a}$ ($J_{z}^{a}\left|
m_{a}\right\rangle =m_{a}\left| m_{a}\right\rangle $) and $J_{z}^{b}$ ($%
J_{z}^{b}\left| m_{b}\right\rangle =m_{b}\left| m_{b}\right\rangle $) with $%
m_{a}=-\frac{N_{a}}{2},...,\frac{N_{a}}{2}$ and $m_{b}=-\frac{N_{b}}{2},...,%
\frac{N_{b}}{2}$.

As a first step we need to diagonalize the unperturbed Hamiltonian (\ref{m28}%
), which can be done by performing the following $\theta
=\frac{\pi }{4}$ rotation on the operators $J_{z}^{a}$,
$J_{z}^{b}$:
\begin{equation}
\begin{array}{cc}
\begin{array}{c}
\overline{O}_{z}^{1}=a_{1}J_{z}^{a}-a_{2}J_{z}^{b} \\
\overline{J}_{z}^{2}=a_{1}J_{z}^{a}+a_{2}J_{z}^{b}
\end{array}
, & a_{1}=a_{2}=\frac{1}{\sqrt{2}},
\end{array}
\label{m30}
\end{equation}
while an analogous rotation needs to be carried out on $J_{x}^{a}$, $%
J_{x}^{b}$ entering the perturbation (\ref{m29}). As a result we
get:
\begin{equation}
\overline{H}=\frac{1}{2}\left( \Lambda -\Lambda _{ab}\right)
\left( \overline{O}_{z}^{1}\right) ^{2}+\frac{1}{2}\left( \Lambda
+\Lambda
_{ab}\right) \left( \overline{J}_{z}^{2}\right) ^{2}-\frac{2K}{\sqrt{2}}%
\overline{J}_{x}^{2},  \label{m31}
\end{equation}
which, by defining $\Lambda _{1}=\Lambda -\Lambda _{ab}$, $\Lambda
_{2}=\Lambda +\Lambda _{ab}$, and $\widehat{O}_{z}^{1}=\frac{\overline{O}%
_{z}^{1}}{\sqrt{2}}$, $\widehat{O}_{x}^{1}=\frac{\overline{O}_{x}^{1}}{\sqrt{%
2}}$, $\widehat{J}_{z}^{2}=\frac{\overline{J}%
_{z}^{2}}{\sqrt{2}}$, $\widehat{J}_{x}^{2}=\frac{\overline{J}_{x}^{2}}{\sqrt{%
2}}$, can be cast in the final form:
\begin{equation}
\widehat{H}=\Lambda _{1}\left( \widehat{O}_{z}^{1}\right)
^{2}+\Lambda _{2}\left( \widehat{J}_{z}^{2}\right)
^{2}-2K\widehat{J}_{x}^{2}. \label{m32}
\end{equation}
In the following Section we will find analytical expressions for
the eigenvalues and the eigenvectors up to second order by
performing perturbation theory in the tunneling amplitude.

\section{Stationary states}

In the present Section we apply second-order perturbation theory to the
Hamiltonian of Eq. (\ref{m32}) in the small tunneling amplitude limit, which
allows us to derive analytical expressions for the stationary states of the
system.

In order to pursue this task let us rewrite Eq. (\ref{m32}) in dimensionless
form by assuming $\frac{\Lambda _{1}}{2}$ as unit of energy:
\begin{equation}
\widehat{H}=2\left( \widehat{O}_{z}^{1}\right) ^{2}+2\lambda \left( \widehat{%
J}_{z}^{2}\right) ^{2}-2k\widehat{J}_{x}^{2},  \label{m33}
\end{equation}
where $\lambda =\frac{\Lambda _{2}}{\Lambda _{1}}$ and $k=\frac{2K}{\Lambda
_{1}}$, then take
\begin{equation}
\widehat{H}_{0}=2\left( \widehat{O}_{z}^{1}\right) ^{2}+2\lambda \left(
\widehat{J}_{z}^{2}\right) ^{2}  \label{m34}
\end{equation}
as unperturbed Hamiltonian and
\begin{equation}
\widehat{H}_{I}=-2k\widehat{J}_{x}^{2},  \label{m35}
\end{equation}
as a small perturbation term. Here $%
\widehat{J}_{i}^{2}$, $i=x,y,z$, obey the usual angular momentum algebra
and the following relation holds:
\begin{equation}
\left( \widehat{J}^{2}\right) ^{2}=\frac{N_{2}}{2}\left( \frac{%
N_{2}}{2}+1\right) ,
\label{m35a}
\end{equation}
where:
\begin{equation}
N_{2}=\frac{N_a+N_b}{2}.
\label{m36}
\end{equation}
In principle, the rotated basis $\left\{ \left| m_{1},m_{2}\right\rangle
\right\} =\left\{ \left| m_{1}=\frac{1}{2}\left( m_{a}-m_{b}\right)
\right\rangle ,\left| m_{2}=\frac{1}{2}\left( m_{a}+m_{b}\right)
\right\rangle \right\} $ of the unperturbed Hamiltonian (\ref{m34}) is given by the eigenvectors of $\widehat{O}_{z}^{1}=\frac{\widehat{J}_{z}^{a}-\widehat{J}_{z}^{b}}{2}
$ ($\widehat{O}_{z}^{1}\left| m_{1}\right\rangle =m_{1}\left|
m_{1}\right\rangle $) and $\widehat{J}_{z}^{2}$ ($\widehat{J}_{z}^{2}\left|
m_{2}\right\rangle =m_{2}\left| m_{2}\right\rangle $) with $m_{1}=-\frac{%
\left|j_a-j_b\right|}{2},...,\frac{\left|j_a-j_b\right|}{2}$ and $m_{2}=-\frac{(j_a+j_b)}{2},...,\frac{(j_a+j_b)}{2%
}$, whose corresponding eigenvalues are $\widehat{E}%
_{m_{1},m_{2}}^{\left( 0\right) }=2\left( m_{1}\right) ^{2}+2\lambda \left(
m_{2}\right) ^{2}$.

The presence of the operator $\widehat{O}_{z}^{1}$, which does not commute with the perturbation
term $\widehat{H}_{I}$, makes the problem of finding eigenvalues and eigenvectors of the full Hamiltonian (\ref{m33}) within perturbation theory much more involved. In order to simplify the treatment and carry out analytical calculations while retaining the relevant phenomenology, we concentrate on the particular case of a
binary mixture where the two species are equally populated, i. e. $%
N_{a}=N_{b}$, and have the same population imbalance between the two wells,
i. e. $m_{a}=m_{b}$. This situation allows us to describe the quantum
dynamics of the system in correspondence of the MQST regime, for
which we need a completely localized initial state.
That fixes $m_{1}=0$ while $m_{2}=m_{a}=-\frac{%
N_{a}}{2},...,\frac{N_{a}}{2}$ could be an even or odd integer depending on $%
N_{a}$ even or odd, and leads to the following zero-order eigenvalues: $%
\widehat{E}_{0,m_{2}^{\pm }}^{\left( 0\right) }=2\lambda \left( m_{2}\right)
^{2}=\frac{\lambda }{2}\left( m_{a}+m_{b}\right) ^{2}$. Each eigenvalue is
two-fold degenerate, with the only exception of the ground state for $N_{a}$
even, $\widehat{E}_{0,0}^{\left( 0\right) }=0$, which is nondegenerate. The
two-dimensional subspace of degeneracy is spanned by the states $\left|
0,\pm m_{2}\right\rangle $ (where $\widehat{J}_{z}^{2}\left| 0,\pm
m_{2}\right\rangle =\pm m_{2}\left| 0,\pm m_{2}\right\rangle $) and the
corresponding zero-order eigenvectors are:
\begin{equation}
\left| \widehat{h}_{0,m_{2}^{\pm }}^{\left( 0\right) }\right\rangle =\left|
0,m_{2}^{\pm }\right\rangle =\frac{1}{\sqrt{2}}\left( \left|
0,m_{2}\right\rangle \pm \left| 0,-m_{2}\right\rangle \right) .  \label{m37}
\end{equation}

By switching on the perturbation term (\ref{m35}) it is possible to show
that the degeneration is lifted starting from the levels with
smaller $m_{2}$; in general
the double degeneracy of the zero-order eigenvalues $\widehat{E}%
_{0,m_{2}^{\pm }}^{\left( 0\right) }$ will be lifted at the $2m_{2}$-th
order of perturbation theory \cite{bishop1}. By applying perturbation
theory \cite{cohen1} up to order $k^{2}$, we obtain the following
corrected eigenvalues:
\begin{equation}
\widehat{E}_{0,m_{2}^{\pm }}^{\left( 2\right) }=2\lambda \left( m_{2}\right)
^{2}+\frac{k^{2}}{\lambda }\frac{j_{2}\left( j_{2}+1\right) +\left(
m_{2}\right) ^{2}}{4\left( m_{2}\right) ^{2}-1};\text{ \ \ \ \ \ \ \ }%
m_{2}\neq 1,\frac{1}{2},  \label{m38}
\end{equation}
\begin{equation}
\widehat{E}_{0,1^{\pm }}^{\left( 2\right) }=2\lambda +\frac{k^{2}}{\lambda }\left( \frac{j_{2}\left( j_{2}+1\right)
+1}{3}\pm \frac{j_{2}\left( j_{2}+1\right) }{2}\right) ;\text{ \ \ \ \ \ \ \
}N_{2}\text{ even,}  \label{m39}
\end{equation}
\begin{equation}
\widehat{E}_{0,\frac{1}{2}^{\pm }}^{\left( 2\right) }=\frac{\lambda }{2}\mp k\sqrt{j_{2}\left( j_{2}+1\right) +%
\frac{1}{4}}-\frac{k^{2}}{4\lambda }\left( j_{2}\left( j_{2}+1\right) -\frac{%
3}{4}\right) ;\text{ \ \ \ \ \ \ \ }N_{2}\text{ odd,}  \label{m40}
\end{equation}
where $j_{2}=\frac{N_{2}}{2}$. Furthermore, for $N_{2}$ even, the
nondegenerate ground state $\left| 0,0\right\rangle $ belongs to the
symmetry class of $\left| 0,m_{2}^{+}\right\rangle $. The corresponding
eigenvectors, up to order $k^{2}$, are given in the Appendix.

In the following Sections we use the analytical expressions of energy
eigenvectors derived in the Appendix, see Eqs. (\ref{m41})-(\ref{m48}), in order to study
the quantum evolution of $%
\left\langle \widehat{J}_{z}^{2}\left( \tau \right) \right\rangle
=\left\langle \frac{1}{2}\left( J_{z}^{a}\left( \tau \right)
+J_{z}^{b}\left( \tau \right) \right) \right\rangle $, that is the number
difference of bosons of species $a$ and $b$ between the two wells of the
potential.

\section{Dynamics: completely localized initial states}

In this Section we investigate the quantum evolution of the number
difference of bosons of species $a$ and $b$ between the two wells assuming a
completely localized state as initial condition. That could be interesting
in order to elucidate the quantum behavior of the system in correspondence
of the classical MQST regime and to put in evidence new phenomena including
quantum coherence in a multicomponent system. In such a case we will study
both the short and the long time dynamics: as a result a rich behaviour
emerges, ranging from small amplitude oscillations and collapses and
revivals to coherent tunneling. Although such a physics is well known for
the single component Bose Josephson junction, in our case the dynamics shows
that the two species can coexist in the same potential well as if there
would be an attractive interaction between them.

As a first step let us recall the general formula which gives the time
evolution of the mean value of $\widehat{J}_{z}^{2}=\frac{1}{2}\left(
J_{z}^{a}+J_{z}^{b}\right) $ \cite{cohen1}:
\begin{equation}
\left\langle \widehat{J}_{z}^{2}\left( \tau \right) \right\rangle
=\sum_{n=m_{2}^{\pm}}\sum_{n^{\prime }=m_{2}^{\pm }}\phi _{n}^{\ast }\phi _{n^{\prime }}\left\langle \widehat{h}_{0,n}\right| \widehat{J}_{z}^{2}\left|
\widehat{h}_{0,n^{\prime }}\right\rangle e^{i\left( \widehat{E}%
_{0,n}-\widehat{E}_{0,n^{\prime }}\right) \tau },  \label{m49}
\end{equation}
where $\tau =\frac{\Lambda _{1}}{2\hbar }t$ is the dimensionless time, the
sums are over all the eigenvectors $\left| \widehat{h}_{0,m_{2}^{\pm }}\right\rangle $, being $m_{2}=0$ or $\frac{1}{2},...,\frac{N_{2}}{2}$, and $\phi _{n}$
are the projections of the initial state $\left| \psi \left( 0\right)
\right\rangle $ on the basis $\left| \widehat{h}_{0,m_{2}^{\pm
}}\right\rangle $:
\begin{equation}
\left| \psi \left( 0\right) \right\rangle =\sum_{n=m_{2}^{\pm }}\phi _{n}\left| \widehat{h}_{0,n}\right\rangle .
\label{m50}
\end{equation}
So it is clear how the knowledge of eigenvalues and eigenvectors is enough
in order to study the quantum evolution of $\widehat{J}_{z}^{2}$, the Bohr
frequencies involved, $\widehat{E}_{0,n}-\widehat{E}_{0,n^{\prime
}}$, and the corresponding weights $\phi _{n}^{\ast }\phi _{n^{\prime }}\left\langle \widehat{h}_{0,n}\right| \widehat{J}_{z}^{2}\left|
\widehat{h}_{0,n^{\prime }}\right\rangle $.

Let us now study the dynamics of the system when all the bosons of species $a
$ and $b$ are initially contained in one of the two wells of the potential,
say the right one, and then the imbalances of the two species coincide, so
that $N_{aR}=N_{a}$, $N_{aL}=0$, $N_{bR}=N_{b}$, $N_{bL}=0$; furthermore the
two species are equally populated, i.e. $N_{a}=N_{b}$. That implies $m_{1}=0$
and $m_{2}=\frac{N_{a}}{2}=\frac{N_{2}}{2}$ in our \textit{center of mass}
rotated basis. The corresponding initial condition is:
\begin{equation}
\left| \psi \left( 0\right) \right\rangle =\left| 0,\frac{N_{2}}{2}%
\right\rangle .  \label{m51}
\end{equation}
In order to investigate the short timescales evolution we need to keep terms
up to second order in the tunneling amplitude $k$ when we compute the
weights in Eq. (\ref{m49}). We find that:
\begin{equation}
\left\langle \left( \widehat{J}_{z}^{2}\right) ^{\left( 2\right) }\left(
\tau \right) \right\rangle =\frac{N_{2}}{2}+\frac{k^{2}N_{2}}{2\lambda
^{2}\left( N_{2}-1\right) ^{2}}\left[ \cos (\omega _{\mu }\tau )-1\right] ,
\label{m52}
\end{equation}
where the frequency involved is:
\begin{equation}
\omega _{\mu }=\widehat{E}_{0,\frac{N_{2}}{2}^{\pm }}^{\left( 2\right) }-%
\widehat{E}_{0,\left( \frac{N_{2}}{2}-1\right) ^{\pm }}^{\left( 2\right)
}=2\lambda \left( N_{2}-1\right) -\frac{k^{2}}{\lambda }\frac{N_{2}+1}{%
\left( N_{2}\right) ^{2}-4N_{2}+3}.  \label{m53}
\end{equation}
At short timescales small amplitude oscillations with frequency $\omega
_{\mu }$ around the initial condition ($N_{2R}=N_{2}$, $N_{2L}=0$) are
observed and that coincides with a strongly self-trapped regime.

\begin{figure}[tbp]
\centering
\includegraphics[scale=0.8]{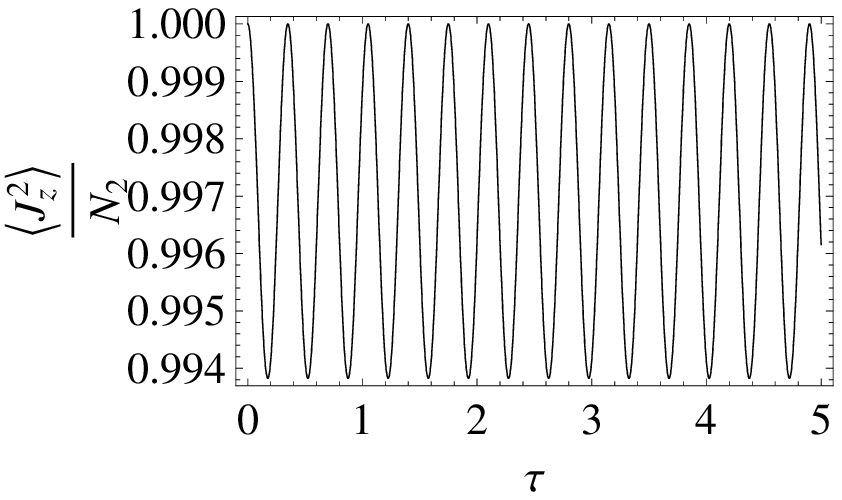}\newline
\includegraphics[scale=0.8]{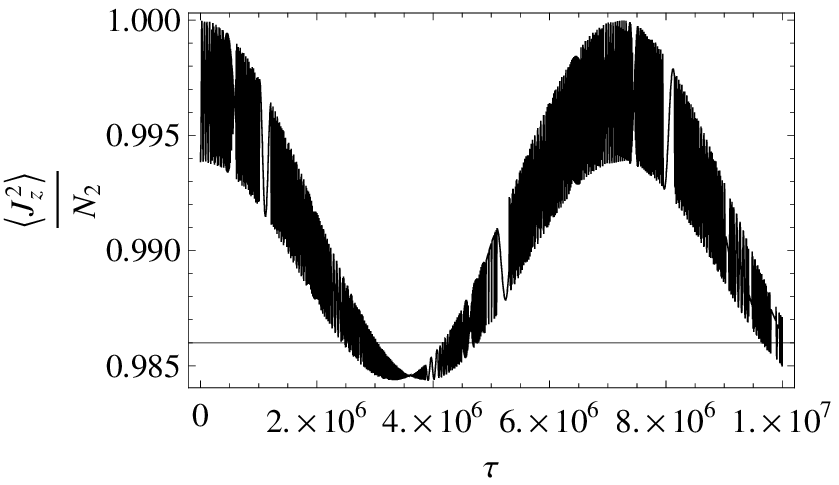}\newline
\includegraphics[scale=0.8]{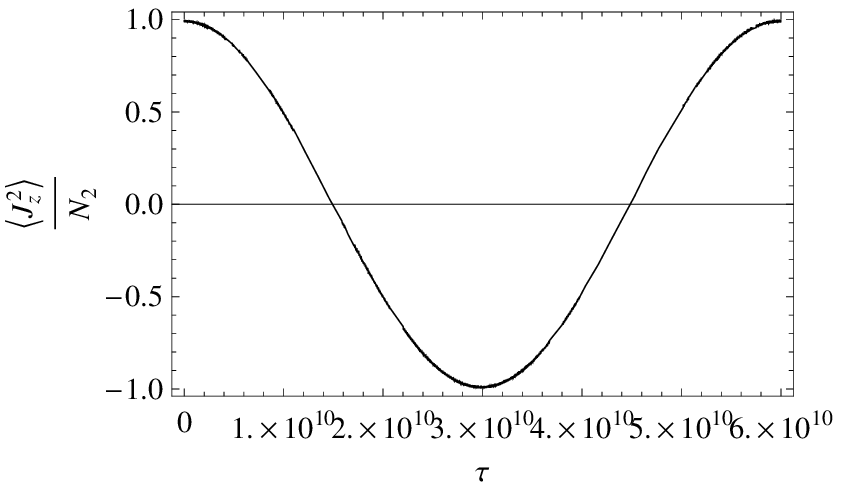}\newline
\caption{(Color online)Time evolution of the relative boson number difference between the
two traps for different timescales. The value of $k$ is $k=0.5$ and the
boson number is $N_2=10.$}
\label{fig:dynamics}
\end{figure}

In order to investigate the dynamics at longer timescales we have to take
into account also the small splittings $\Delta \widehat{E}_{0,\frac{N_{2}}{2}%
^{\pm }}$ and $\Delta \widehat{E}_{0,\left( \frac{N_{2}}{2}-1\right) ^{\pm }}
$ of the two higher pairs of quasidegenerate eigenvalues which provide two
further frequencies (see Ref. \cite{bishop2} for the derivation):
\begin{equation}
\begin{array}{c}
\omega _{0}=\Delta \widehat{E}_{0,\frac{N_{2}}{2}^{\pm }}=\frac{k^{N_{2}}}{%
\lambda ^{N_{2}-1}}\frac{N_{2}}{2^{N_{2}-2}\left( N_{2}-1\right) !} \\
\omega _{1}=\Delta \widehat{E}_{0,\left( \frac{N_{2}}{2}-1\right) ^{\pm }}=%
\frac{k^{N_{2}-2}}{\lambda ^{N_{2}-3}}\frac{\left( N_{2}-1\right) \left(
N_{2}-2\right) }{2^{N_{2}-4}\left( N_{2}-3\right) !}
\end{array}
.  \label{m54}
\end{equation}
The whole result is:
\begin{eqnarray}
\left\langle \left( \widehat{J}_{z}^{2}\right) ^{\left( 2\right) }\left(
\tau \right) \right\rangle  &=&\frac{N_{2}}{2}\cos (\omega _{0}\tau )+\frac{%
k^{2}N_{2}}{4\lambda ^{2}\left( N_{2}-1\right) ^{2}}\left[ \frac{N_{2}}{2}%
\left[ \cos (\omega _{1}\tau )-\cos (\omega _{0}\tau )\right] \right.
\nonumber \\
&&\left. +2\cos (\omega _{\mu }\tau )\cos (\frac{\omega _{1}}{2}\tau )-\cos
(\omega _{1}\tau )-\cos (\omega _{0}\tau )\right] ,  \label{m55}
\end{eqnarray}
and, by putting $\omega _{0}=\omega _{1}=0$, the short timescale dynamics,
Eq. (\ref{m52}), is recovered. Summarizing, at longer timescales the two
species bosons are still localized in the initial potential well but the
quantum dynamics exhibits collapses and complete revivals. Indeed the
coefficient $\cos \left( \frac{\omega _{1}}{2}\tau \right) $, which
multiplies the higher frequency term $\cos \left( \omega _{\mu }\tau \right)
$, gives rise to the beat, which is responsible for the observed collapses
and revivals at timescales fixed by $\omega _{1}$, as shown in Fig. \ref
{fig:dynamics}. Finally, at very large timescales determined by the
frequency $\omega _{0}$ all the bosons tunnel coherently back and forth
between the two traps; only the first term $\cos (\omega _{0}\tau )$ is
responsible of such a coherent tunneling, since all harmonic functions
containing the frequency $\omega _{1}$ and $\omega _{\mu }$ are small in
amplitude and proportional to $k^{2}$, thus they are unable to transfer
bosons from one trap to the other.

The tunneling dynamics within macroscopic quantum self-trapping regime
described above is analogous to that of the $\pi$-mode fixed point obtained
by the Gross-Pitaevski approach \cite{mix3}, where the two species localize
in the same well despite the repulsive interaction between them. Let us
finally note that, despite the explicit dependence on $\lambda$ of the
frequencies (\ref{m53})-(\ref{m54}), the different physics related to the
three time scales described above is simply due to the energy splitting
introduced by the renormalized tunneling for small $\Lambda_{ab}$. Thus in
the case of a mixture of BECs with equal population the dynamics remains
similar to that of a single component BEC, apart the coexistence of the two
species in the same well.

As for the experimental detection of the long timescales phenomena
(collapses/revivals and coherent tunneling), since the time for their
appearance is abruptly increased with $N_{2}$, this implies a rapid decrease
of the characteristic frequencies rendering more difficult the observation
of the intermediate and long time behavior in current BEC experiments.
Indeed pure condensates consisting of $1150\pm150$ atoms of $^{87}Rb$ loaded
in a double well have been recently realized \cite{bec5}\cite{gati1} thus
rendering the detection of the intermediate time behavior possible. Mixtures
with a number of atoms ranging from $9\times10^3$ and $5\times10^3$ ($^{87}Rb
$ and $^{41}K$ \cite{bec6}) to $4\times10^4$ and $9\times10^4$ ($^{85}Rb$
and $^{87}Rb$ \cite{bec7}) have also been recently realized, but in this
case very small characteristic frequencies are implied. However, these
phenomena may be relevant for molecular systems where the number of
vibrational excited quanta is small.

%RC

In the next Section we will further investigate the dynamics of the system
by assuming as initial state a simple coherent state and then study the
formation of a particular superposition of such coherent states, the so
called Schroedinger cat states.

\section{Dynamics: coherent spin initial states and Schroedinger cat states}

In this Section we choose as initial condition a simple coherent spin state \cite{arecchi}
and study the short-time scale evolution of number difference; in this way a
more complex dynamics will appear. Finally, we study the generation of
Schroedinger cat states; in particular, we focus on the contrast in the
momentum distribution and show how it vanishes for a two-component cat state.

\begin{figure}[tbp]
\centering
\includegraphics[scale=0.9]{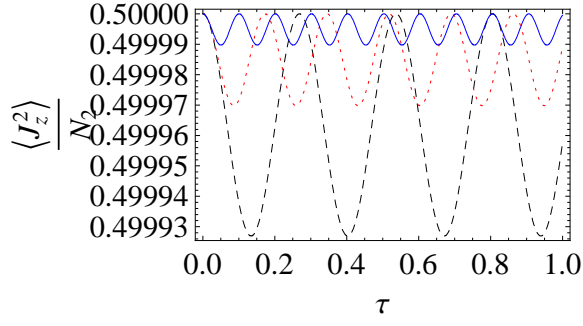}\newline
\caption{(Color online)Time evolution of the relative boson number difference between the
two traps for different boson numbers. The value of $k$ is $k=0.1,\protect%
\lambda=1.3$ and $N_2=10.$(black-dashed line), $N_2=15.$(red-dotted line), $%
N_2=25.$(blue-straight line), while $\protect\theta=\protect\pi/2$ and $%
\protect\phi=\protect\pi/4$.}
\label{fig:st-dynamics}
\end{figure}

\begin{figure}[tbp]
\centering
\includegraphics[scale=0.9]{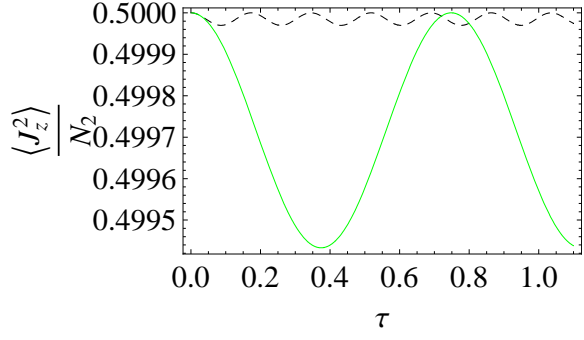}\newline
\caption{(Color online)Time evolution of the relative boson number difference between the
two traps for $N_2=10$. The value of $k$ is $k=0.1$ while $\protect\lambda%
=1.3$(black-dashed line) and $\protect\lambda=0.3$(green-straight line),
while $\protect\theta=\protect\pi/2$ and $\protect\phi=\protect\pi/4$.}
\label{fig:st-dynamics-lambda}
\end{figure}

Let us start by considering as initial condition the following coherent spin
state \cite{arecchi}:
\begin{equation}
\left| \psi \left( 0\right) \right\rangle =C\sum_{m_{2}=-N_{2}/2}^{N_{2}/2}%
\sqrt{\frac{N_{2}!}{\left( \frac{N_{2}}{2}+m_{2}\right) !\left( \frac{N_{2}}{%
2}-m_{2}\right) !}}\tan ^{m_{2}}\left( \frac{\theta }{2}\right)
e^{-im_{2}\phi }\left| 0,m_{2}\right\rangle ,  \label{m56}
\end{equation}
where the coefficient $C$ is:
\begin{equation}
C=\sin ^{N_{2}/2}\left( \frac{\theta }{2}\right) \cos ^{N_{2}/2}\left( \frac{%
\theta }{2}\right) e^{-i\left( N_{2}/2\right) \phi },  \label{m57}
\end{equation}
and $\theta $ and $\phi $ are two angles characterizing the superposition.
The time evolution of the mean value of $\widehat{J}_{z}^{2}$ up to first
order in the tunneling amplitude $k$ is given by:
\begin{eqnarray}
\left\langle \left( \widehat{J}_{z}^{2}\right) ^{\left( 1\right) }\left(
\tau \right) \right\rangle &=&\left\langle \left( \widehat{J}_{z}^{2}\right)
^{\left( 0\right) }\left( \tau \right) \right\rangle +\frac{k}{\lambda}%
\left( \frac{\sin (\theta )}{2}\right) ^{N_{2}}\left[ C_{1}\left[ \cos
(\omega _{e}\tau )-1\right] +C_{2}\sin \left( \omega _{e}\tau \right) \right.
\nonumber \\
&&\left. +\sum_{n=0or1/2}^{N_{2}/2-1}\frac{N_{2}!}{\left( \frac{N_{2}}{2}%
+n\right) !\left( \frac{N_{2}}{2}-n\right) !}\frac{N_{2}-2n}{2\left(
2n+1\right) }A_{n}\right] ,  \label{m58}
\end{eqnarray}
where
\begin{equation}
A_{n}=\tan ^{2n+1}\left( \frac{\theta }{2}\right) \left[ \cos \left(
F_{n}\tau +\phi \right) -\cos \left( \phi \right) \right] -\frac{1}{\tan
^{2n+1}\left( \frac{\theta }{2}\right) }\left[ \cos \left( F_{n}\tau -\phi
\right) -\cos \left( \phi \right) \right]  \label{m59}
\end{equation}
with frequencies $F_{n}=\widehat{E}_{0,\left( n+1\right) ^{\pm }}^{\left(
0\right) }-\widehat{E}_{0,\left( n\right) ^{\mp }}^{\left( 0\right)
}=\lambda\left(4n+2\right)$. Furthermore the coefficients $C_{1}$ and $C_{2}$
are given by:
\begin{equation}
C_{1}=\frac{N_{2}!}{\left( \frac{N_{2}}{2}+1\right) !\left( \frac{N_{2}}{2}%
-1\right) !}\cos \left( \phi \right) \left\{ \left( \frac{N_{2}}{6}-\frac{1}{%
3}\right) \left[ \tan ^{3}\left( \frac{\theta }{2}\right) -\frac{1}{\tan
^{3}\left( \frac{\theta }{2}\right) }\right] -\left( \frac{N_{2}}{2}%
+1\right) \left[ \tan \left( \frac{\theta }{2}\right) -\frac{1}{\tan \left(
\frac{\theta }{2}\right) }\right] \right\} ,  \label{m60}
\end{equation}
\begin{equation}
C_{2}=\frac{N_{2}!}{\left( \frac{N_{2}}{2}+1\right) !\left( \frac{N_{2}}{2}%
-1\right) !}\left[ \tan \left( \frac{\theta }{2}\right) +\frac{1}{\tan
\left( \frac{\theta }{2}\right) }\right] \left[ \left( \frac{N_{2}}{6}-\frac{%
1}{3}\right) \sin \left( 3\phi \right) -\left( \frac{N_{2}}{2}+1\right) \sin
\left( \phi \right) \right] ,  \label{m61}
\end{equation}
for $N_{2}$ even, and
\begin{equation}
C_{1}=\frac{N_{2}!}{\left( \frac{N_{2}}{2}+\frac{1}{2}\right) !\left( \frac{%
N_{2}}{2}-\frac{1}{2}\right) !}\frac{N_{2}-1}{8}\cos \left( \phi \right) %
\left[ \tan ^{2}\left( \frac{\theta }{2}\right) -\frac{1}{\tan ^{2}\left(
\frac{\theta }{2}\right) }\right] ,  \label{m62}
\end{equation}
\begin{equation}
C_{2}=\frac{N_{2}!}{\left( \frac{N_{2}}{2}+\frac{1}{2}\right) !\left( \frac{%
N_{2}}{2}-\frac{1}{2}\right) !}\frac{N_{2}-1}{8}\sin \left( 2\phi \right) %
\left[ \tan \left( \frac{\theta }{2}\right) +\frac{1}{\tan \left( \frac{%
\theta }{2}\right) }\right] ,  \label{m63}
\end{equation}
for $N_{2}$ odd, respectively. Finally, for $N_{2}$ even, the zero-order
mean value $\left\langle \left( \widehat{J}_{z}^{2}\right) ^{\left( 0\right)
}\left( \tau \right) \right\rangle $ is given by:
\begin{eqnarray}
\left\langle \left( \widehat{J}_{z}^{2}\right) ^{\left( 0\right) }\left(
\tau \right) \right\rangle &=&-\frac{N_{2}}{2}\cos \left( \theta \right)
+\left( \frac{\sin (\theta )}{2}\right) ^{N_{2}}\frac{N_{2}!}{\left( \frac{%
N_{2}}{2}+1\right) !\left( \frac{N_{2}}{2}-1\right) !}  \nonumber \\
&&\left\{ \left[ \tan ^{2}\left( \frac{\theta }{2}\right) -\frac{1}{\tan
^{2}\left( \frac{\theta }{2}\right) }\right] \left[ \cos (\omega _{e}\tau )-1%
\right] +2\sin \left( 2\phi \right) \sin \left( \omega _{e}\tau \right)
\right\} ,  \label{m64}
\end{eqnarray}
where the dominant frequency $\omega _{e}$ is equal to $\omega _{e}=\widehat{%
E}_{0,1^{+}}^{\left( 2\right) }-\widehat{E}_{0,1^{-}}^{\left( 2\right) }=%
\frac{k^{2}}{\lambda}\frac{N_{2}}{2}\left( \frac{N_{2}}{2}+1\right) $. The
corresponding expression for $N_{2}$ odd is:
\begin{eqnarray}
\left\langle \left( \widehat{J}_{z}^{2}\right) ^{\left( 0\right) }\left(
\tau \right) \right\rangle &=&-\frac{N_{2}}{2}\cos \left( \theta \right) +%
\frac{1}{2}\left( \frac{\sin (\theta )}{2}\right) ^{N_{2}}\frac{N_{2}!}{%
\left( \frac{N_{2}}{2}+\frac{1}{2}\right) !\left( \frac{N_{2}}{2}-\frac{1}{2}%
\right) !}  \nonumber \\
&&\left\{ \left[ \tan \left( \frac{\theta }{2}\right) -\frac{1}{\tan \left(
\frac{\theta }{2}\right) }\right] \left[ \cos (\omega _{e}\tau )-1\right]
-2\sin \left( \phi \right) \sin \left( \omega _{e}\tau \right) \right\} ,
\label{m65}
\end{eqnarray}
where $\omega _{e}=\widehat{E}_{0,\frac{1}{2}^{-}}^{\left( 2\right) }-%
\widehat{E}_{0,\frac{1}{2}^{+}}^{\left( 2\right) }=2k\sqrt{\frac{N_{2}}{2}%
\left( \frac{N_{2}}{2}+1\right) +\frac{1}{4}}$. As one can see, the dominant frequency is gradually suppressed with the number of bosons $N_{2}=\frac{1}{2}\left( N_{a}+N_{b}\right)$ \cite{bishop1}. This
is clearly seen in Fig. \ref{fig:st-dynamics} where the boson number
difference between the two traps is plotted for different $N_2$ (even)
values as a function of the dimensionless time $\tau$. One also notices a decrease of the oscillation amplitude at
increasing $N_2$. \newline
The effect of $\lambda$ is instead shown in Fig. \ref{fig:st-dynamics-lambda}
where the short-time dynamics of the boson number difference is analyzed for
two values of the interspecies interaction. When $\lambda$ increases the
amplitude of the oscillations decreases. The detection of the mixture
dynamics is thus more favorable for values of $\lambda$ smaller than unity.

.

.

\subsection{Cat states}

\begin{figure}[tbp]
\centering
\includegraphics[scale=0.9]{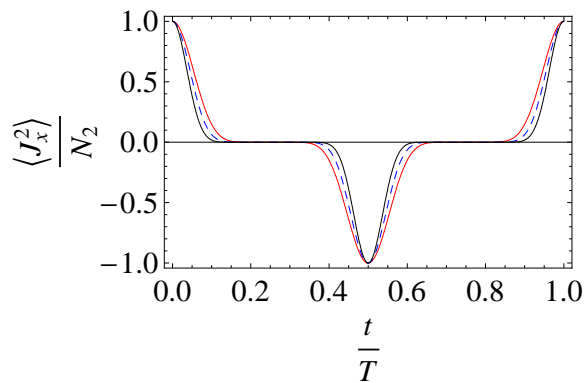}
\caption{(Color online)Contrast in the time-evolution of $\langle J_x^2\rangle$ for $%
\protect\theta=\protect\pi/2$ and $\protect\phi=\protect\pi/4$ and even
number of bosons. The red line is for $N_2=10.$, the blue one for $N_2=14.$
and the black one for $N_2=20.$. The interval in which the contrast is zero
increases with increasing $N_2$. }
\label{fig:contrast}
\end{figure}

Let us consider the coherent spin state (\ref{m56}); the expectation value
of the Hamiltonian (\ref{m33}) on such state is given by:
\begin{equation}  \label{eq:exp-value}
\langle \psi \left( 0\right) | \widehat{H} \left| \psi \left( 0\right)
\right \rangle=2\lambda n^2/2-2k \sqrt{(N_2/2)^2-n^2}\cos \phi,
\end{equation}
where $n=-(N_2/2)\cos \theta$ and has the maximum value for $%
\phi=0,\theta=\pi/2$. This result also corresponds to the mean-field result
for the energy. Now, starting from the coherent spin state (\ref{m56}) we
are interested in looking for Schroedinger cat states. Such states are
quantum superposition of macroscopic states and their realization has
already been suggested for a single species Bose-Josephson junction in \cite
{minguzzi1}. Also in the case of a Bose-Josephson junction with binary
mixtures one might realize cat states from the time-evolution of an
initially coherent state following a sudden rise of the barrier between the
two wells. Thus we consider at time $t=0^+$ a zero inter-well coupling $k$,
i.e. the time evolution is governed by the Hamiltonian $H_0$ in Eq. (\ref
{m34}). For each basis vector $|0,m_2\rangle$ of the coherent state (\ref
{m56}), the time-evolution is given by $|0,m_2\rangle(t)= e^{-i 2\pi m_2^2
t/T}|0,m_2\rangle$, where $T=\hbar \pi/\lambda$ is the so-called revival
time such that $|\psi(T)\rangle=|\psi(0)\rangle$. Considering now the times $%
T/2p$, $p$ integer, the time evolution of the coherent state is governed by
the factor $\exp( -i\pi m_2^2/p)$ which satisfies the property $\exp( -i\pi
(m_2+p)^2/p)=(-1)^p \exp( -i\pi m_2^2/p)$, depending on the parity of $p$.
For the choice of even $p$, a discrete Fourier transform leads to the cat
state:
\begin{equation}
|\psi(T/2p)\rangle=\sum_{k=0}^{p-1}u_k e^{i\pi k N_2/p}|e^{-i2\pi k/p }\psi
\rangle,
\end{equation}
i.e. a superposition of $p$ coherent states, where $u_k=1/p
\sum_{m_2=0}^{p-1}e^{-i\pi m_2^2/p}e^{i2\pi km/p}$. In particular, the cat
state affects the momentum distribution. This dependence could be important
to probe experimentally their existence. In particular, when considering the
two-component cat state, i.e. for the choice $p=2$, one obtains that the
contrast in the momentum distribution, i.e. the expectation value of $%
\widehat{J}_x^2$ on the unperturbed state, vanishes\cite{minguzzi1}.
Furthermore, the amplitude of the intervals of time in which the contrast is
zero increases with increasing $N_{2}$ as clearly shown in Fig. \ref
{fig:contrast}.

It should be noted that despite the close similarity in the behavior of the
contrast between the single component BJJ and the double one, the mixture
will be a better candidate for the creation and detection of cat states. In
fact their creation time is $\pi \hbar/\lambda$ and since for repulsive
interaction between the two species and $\Lambda>\Lambda_{ab}$ we get $%
\lambda=-[(1+\Lambda/\Lambda_{ab})/(1-\Lambda/\Lambda_{ab})]>1$, such time
can be made short enough to render their detection more favorable. For
example by fixing the ratio of $^{87}Rb-^{87}Rb$ interaction to $%
^{87}Rb-^{85}Rb$ interaction to be 2.13, a parameter accessible in the JILA
setup\cite{bec6}, the detection time is twice smaller than the case of a
single component BEC.

\section{Conclusions and perspectives}

In this paper we investigated the quantum dynamics of a Bose
Josephson junction made of a binary mixture of BECs loaded in a
double well potential within the two-mode approximation. We
focused on the small tunneling amplitude limit and adopted the
angular momentum representation for the Bose-Hubbard dimer
Hamiltonian. Perturbation theory up to second order in the
tunneling amplitude enabled us perform analytical calculations in
the symmetric case where $\Lambda _{a}=\Lambda _{b}=\Lambda $ and
$K_{a}=K_{b}=K$. In this
way we obtained the energy eigenvalues and eigenstates, whose
knowledge is mandatory in order to investigate the quantum
evolution of the number difference of bosons between the two
potential wells. In order to study the quantum dynamics more easily and analitycally, we restricted to the case in which the two species are equally populated and imposed the condition of equal population imbalance of the species $a$ and $b$ between the two wells. We concentrated on the two following initial
conditions: completely localized states and coherent spin states,
and found a rich and complex behaviour, ranging from small
amplitude oscillations and collapses and revivals to coherent
tunneling. Finally, we
considered the generation of Schroedinger cat states and pointed
out their influence on the momentum distribution through the
vanishing of the contrast. We showed that the creation time can be
rendered short enough in the case of a mixture in order to render
their detection more favorable. That could be crucial in order to
build up an experimental protocol to produce and detect cat states
within such systems.

We stress that in this work we have chosen to study the symmetric
case. This allowed us to obtain analytical results, while giving
rise to the relevant phenomenology which characterizes the physics
of the junction. The general case of different couplings between
the two bosonic species and/or different populations needs to resort to numerical calculations
and will be the subject of a future publication \cite{noi1}.
Another interesting issue which deserves further investigation is
a careful analysis of the quantum manifestations of the
self-trapping transition and in general of the MQST phenomenon in
this more general context.

The complex dynamics of the generalized Bose Josephson junctions
investigated in the present paper could be experimentally testable
within the current technology. For instance, the JILA group
recently \cite{bec6} succeeded in producing a mixture of $^{85}Rb$
and $^{87}Rb$ atoms, whose interactions are widely tunable via
Feshbach resonances. In particular it is possible to fix the
scattering length of $^{87}Rb$ as well as the interspecies one and
to tune the scattering length of $^{85}Rb$. That allows one to
explore the parameter space in a wide range and also to realize
the symmetric regime $\Lambda _{a}=\Lambda _{b}=\Lambda $. Because
of the high degree of experimental control, such a setup could be
employed to reproduce the phenomenology described in this work.

\begin{acknowledgments}
The authors would like to thank M. Salerno for driving their
attention on the topic of Bose Josephson junctions and E. Orignac and A. Minguzzi for discussions and for a critical reading of the manuscript.
\end{acknowledgments}

\section*{Appendix: Order $k^2$ eigenvectors}

The eigenvectors of the full Hamiltonian (\ref{m33}), up to order $k^{2}$, are:
\begin{equation}
\left| \widehat{h}_{0,0}^{\left( 2\right) }\right\rangle =\left( 1-\frac{%
k^{2}}{\lambda ^{2}}\frac{j_{2}\left( j_{2}+1\right) }{4}\right) \left|
0,0\right\rangle +\frac{k}{\lambda }\sqrt{\frac{j_{2}\left( j_{2}+1\right) }{%
2}}\left| 0,1^{+}\right\rangle +\frac{k^{2}}{8\lambda ^{2}}\sqrt{\frac{%
j_{2}\left( j_{2}+1\right) \left[ j_{2}\left( j_{2}+1\right) -2\right] }{2}}%
\left| 0,2^{+}\right\rangle ,  \label{m41}
\end{equation}
\begin{eqnarray}
\left| \widehat{h}_{0,1^{-}}^{\left( 2\right) }\right\rangle
&=&\left( 1-\frac{k^{2}}{72\lambda ^{2}}\left[ j_{2}\left( j_{2}+1\right) -2%
\right] \right) \left| 0,1^{-}\right\rangle +\frac{k}{6\lambda }%
\sqrt{\left[ j_{2}\left( j_{2}+1\right) -2\right] }\left| 0,2^{-}\right\rangle   \nonumber \\
&&+\frac{k^{2}}{96\lambda ^{2}}\sqrt{\frac{\left[ j_{2}\left( j_{2}+1\right)
-2\right] \left[ j_{2}\left( j_{2}+1\right) -6\right] }{2}}\left| 0,3^{-}\right\rangle ,  \label{m42}
\end{eqnarray}
\begin{eqnarray}
\left| \widehat{h}_{0,1^{+}}^{\left( 2\right) }\right\rangle
&=&\left( 1-\frac{k^{2}}{72\lambda ^{2}}\left[ 19j_{2}\left( j_{2}+1\right)
-2\right] \right) \left| 0,1^{+}\right\rangle -\frac{k}{\lambda }%
\sqrt{\frac{j_{2}\left( j_{2}+1\right) }{2}}\left| 0,0\right\rangle +\frac{k}{6\lambda }\sqrt{\left[ j_{2}\left( j_{2}+1\right)
-2\right] }\left| 0,2^{+}\right\rangle   \nonumber\\
&&+\frac{k^{2}}{96\lambda ^{2}}\sqrt{\frac{\left[ j_{2}\left( j_{2}+1\right)
-2\right] \left[ j_{2}\left( j_{2}+1\right) -6\right] }{2}}\left| 0,3^{+}\right\rangle ,  \label{m43}
\end{eqnarray}
\begin{eqnarray}
\left| \widehat{h}_{0,\frac{1}{2}^{\pm }}^{\left( 2\right)
}\right\rangle  &=&\left( 1-\frac{k^{2}}{32\lambda ^{2}}\left[ j_{2}\left(
j_{2}+1\right) -\frac{3}{4}\right] \right) \left| 0,\frac{1}{2}%
^{\pm }\right\rangle +\frac{k^{2}}{48\lambda ^{2}}\sqrt{\left[ j_{2}\left(
j_{2}+1\right) -\frac{3}{4}\right] \left[ j_{2}\left( j_{2}+1\right) -\frac{%
15}{4}\right] }\left| 0,\frac{5}{2}^{\pm }\right\rangle   \nonumber
\\
&&+\left[ \frac{k}{4\lambda }\sqrt{j_{2}\left( j_{2}+1\right) -\frac{3}{4}}%
\mp \frac{k^{2}}{16\lambda ^{2}}\sqrt{\left[ j_{2}\left( j_{2}+1\right) +%
\frac{1}{4}\right] \left[ j_{2}\left( j_{2}+1\right) -\frac{3}{4}\right] }%
\right] \left| 0,\frac{3}{2}^{\pm }\right\rangle ,  \label{m44}
\end{eqnarray}
\begin{eqnarray}
\left| \widehat{h}_{0,m_{2}^{\pm }}^{\left( 2\right)
}\right\rangle  &=&A_{m_{2}}\left| 0,m_{2}^{\pm }\right\rangle
+B_{m_{2}}^{+}\left| 0,\left( m_{2}+1\right) ^{\pm }\right\rangle
+B_{m_{2}}^{-}\left| 0,\left( m_{2}-1\right) ^{\pm }\right\rangle
+C_{m_{2}}^{+}\left| 0,\left( m_{2}+2\right) ^{\pm }\right\rangle
\nonumber \\
&&+C_{m_{2}}^{-}\left| 0,\left( m_{2}-2\right) ^{\pm
}\right\rangle ;\text{ \ \ \ \ \ \ \ }m_{2}\neq 0,1,\frac{1}{2},  \label{m45}
\end{eqnarray}
where the coefficients are defined as:
\begin{equation}
A_{m_{2}}=1-\frac{k^{2}}{4\lambda ^{2}}\frac{4j_{2}\left( j_{2}+1\right)
\left( m_{2}\right) ^{2}+j_{2}\left( j_{2}+1\right) -4\left( m_{2}\right)
^{4}+3\left( m_{2}\right) ^{2}}{\left( 4\left( m_{2}\right) ^{2}-1\right)
^{2}},  \label{m46}
\end{equation}
\begin{equation}
B_{m_{2}}^{\pm }=\pm \frac{k}{2\lambda }\frac{\sqrt{j_{2}\left(
j_{2}+1\right) -m_{2}\left( m_{2}\pm 1\right) }}{\left( 2m_{2}\pm 1\right) },
\label{m47}
\end{equation}
\begin{equation}
C_{m_{2}}^{\pm }=\frac{k^{2}}{16\lambda ^{2}}\frac{\sqrt{j_{2}\left(
j_{2}+1\right) -m_{2}\left( m_{2}\pm 1\right) }\sqrt{j_{2}\left(
j_{2}+1\right) -\left( m_{2}\pm 1\right) \left( m_{2}\pm 2\right) }}{\left(
m_{2}\pm 1\right) \left( 2m_{2}\pm 1\right) }.  \label{m48}
\end{equation}

\end{document}